\documentstyle[aps,eqsecnum,prbbib,12pt]{revtex} 
\begin{document} 
\title{Chern-Simons Superconductivity at finite magnetic
field} 
\author{Sudhansu S. Mandal $^1$\footnote{electronic address:
ssman@iitk.ernet.in}, S. Ramaswamy $^2$\footnote{electronic address:
suresh@mri.ernet.in}, V. Ravishankar $^3$\footnote{electronic address:
vravi@iitk.ernet.in}\\}

\maketitle
\vspace{0.15in}
\begin{center}
    {\bf Abstract}
\end{center}
\begin{abstract}
We study Chern-Simons (CS) superconductivity in the presence of uniform
external magnetic field of {\it arbitrary strength} for a system of
fermions in two spatial dimensions, which are minimally
coupled both to the CS
and Maxwell gauge fields. We have carried out the computation within
the mean field ansatz. Analysing only the mean field (i.e.,
ignoring the fluctuations of the gauge fields), we find that chemical
potential, susceptibility and magnetization show discontinuities for
integer number of filled Landau levels. Taking into account the
 fluctuations of the gauge fields,
 we find that the masses of the excitations increase with the
magnetic field, and that the
presence of nonlinear magnetic susceptibilities show
the absence of any critical or pseudo critical magnetic field.
Finally, an interesting result is that, unlike
ordinary superconductors, the system is magnetically asymmetric.
\end{abstract}

\vspace{0.3in}
\newpage

\section{{\bf INTRODUCTION}}

The relevance of Chern-Simons (CS) gauge theory of planar phenomena
like quantum Hall effect is by now well established. \cite{qhe}
It has also been recognized that the CS interaction can lead to
a novel kind of superconductivity -- characterized by Parity and
Time reversal (${\cal P}$, ${\cal T}$) violation, no Cooper pair
formation, two penetration depths in Meissner effect, and finally, an
antisymmetric (super) conductivity tensor. Proposed originally as a
model for high-$T_{\mbox{c}}$ superconductors, it attracts continued
theoretical interest, partly due to the novelty of the mechanism
and partly due to the distinct possibility of the
existence of such real systems.

This paper is devoted to a study of CS superconductivity (CSS) in
uniform finite (external) magnetic field. Recall that the conventional
superconducting phase gets destroyed beyond both a critical temperature
and a critical field. While there have been extensive studies of CSS
at finite temperatures ($T$), there is not much work at finite magnetic
field ($B$). We intend to fill this gap here.

CSS was first established at $T=0$ in the pioneering works
of Laughlin \cite{laugh} and Chen, Wilczek, Witten and Halperin,
\cite{chen} who considered spinless fermions.
This was followed by an extension to spin $1/2$ by Hosotani and
Chakravarty \cite{hos} and Chakraborty, Ramaswamy and Ravishankar.
\cite{chak}
The latter found a unique possibility for the existence of CSS
with a ferromagnetic ground state. There exists also an extensive
literature on the $T\neq 0$ properties of CSS. See Ref.~6 for details
and for references to earlier works.
To put it concisely, it was found that CSS gives
a normal insulating nonmagnetic state beyond a certain temperature.
The transition to the normal state is over a rather narrow range
of temperatures, but does not appear sharp enough to qualify
unambiguously to be a phase transition. Thus it is not clear whether
we have at hand a critical or a pseudo critical temperature.

Based on a mean field (MF) analysis, Hetric, Hosotani and Lee
\cite{het1}
conclude that there is also a pseudo critical magnetic field beyond
which CSS would not survive. We propose to study this in detail
by computing explicitly the one loop effective action for the system
in a background magnetic field. Further there is the interesting
question of the system's response when the sign of $B$ is flipped.
We anticipate the system to be magnetically asymmetric around $B=0$.
The interesting region around $B=-b$, ($b$ being the mean
CS magnetic field), where the particles exhibit net zero mean field
will also be examined here.

Finally, we remark here that the study of CSS here and elsewhere
relies on the MF ansatz plus perturbative one loop correction. The
MF ansatz requires justification, which has been attempted in Ref.~6.
Alternatively, one might appeal to the success of the MF picture in
fractional quantum Hall effect (FQHE), \cite{jain} which is
in agreement with experiment. \cite{tsui} We return to a discussion
of its validity here as well, although very briefly and only
contextually.

The paper is organized as follows. Sec. IIA displays
briefly the formalism of the MF ansatz.
Mean field results are presented in Sec. IIB. In Sec. IIC,
the form factors are evaluated.
The effective
Lagrangian for the magnetic field is obtained by the
integration over the fluctuating part of the gauge fields
in sec. IIIA. The behaviour of low-lying excitations are
then discussed in Sec. IIIB. In sec. IV,
we compute nonlinear magnetic susceptibilities
and conclude the paper in Sec. V.

\section{{\bf MEAN FIELD THEORY}}

\subsection{{\bf Formalism }}

Consider a system of non-relativistic spinless fermions in
(2+1) dimensions whose dynamics is governed by the Lagrangian
\begin{equation}
{\cal L} = -\frac{1}{4}F_{\mu\nu}F^{\mu\nu}
    -\frac{\nu}{2}\epsilon^{\mu\nu\lambda}
       a_\mu \partial_\nu a_\lambda
      +\psi^\dagger iD_0\psi-\frac{1}{2m}\left| D_k\psi\right| ^2
      +\psi^{\dagger} \mu \psi -eA_0\rho    \; ,
\label{eq1}
\end{equation}
 where $A(a)$ denotes the Maxwell(CS) gauge field and
the covariant derivative $D_\mu=\partial_\mu -ie(A_\mu +a_\mu)$.
The condition of a fixed density of the fermions is implemented
by introducing the chemical potential $\mu$ (note that we use $\mu$
as a space time index as well; this should cause no confusion).
The last term represents the background neutralizing `classical'
charge density.

We employ the path integral formalism to evaluate the partition
function
\begin{equation}
{\cal Z} = \int\, [dA][da][d\psi][d\psi^{\dagger}]\, e^{i\int \,
           d^3x \, {\cal L}} \; .
 \label{eq2}
\end{equation}
We proceed with the evaluation of ${\cal Z}$ by performing the
integration over fermionic field first. This gives the effective
action for the gauge fields incorporating the accumulated effect of fermions
on the system. The standard method of evaluation
of ${\cal Z}$ is the MF ansatz
in which one smears out the CS magnetic field to obtain a uniform
background (in which the particles move). At $T=0$, this approach can be
justified for large $N$ ($N$ being related to the CS coefficient
$\nu = Ne^2/2\pi $) and for the parameters used in our analysis.
\cite{chen,ssman1} In this case the external magnetic field will
have to also be included along with the mean CS magnetic field.
The effect of the external magnetic field would depend on its direction
relative to the CS field.
We expand the gauge fields around the configuration
\begin{equation}
A_0=A_2=0\; ;\; a_0=a_2=0\; ;\; A_1=A_1^{\mbox{ex}}=-Bx_2
\; ;\; a_1=-bx_2 \; .
 \label{eq3}
 \end{equation}

Keeping terms up to second order in fluctuations, we find
\begin{equation}
{\cal Z}={\cal Z}_{\mbox{MF}}\, \int\, [dA][da]\, e^{i \; S} \; ,
 \label{eq4}
\end{equation}
where the mean field action
\begin{equation}
S_{\mbox{MF}}=-i \,\ln {\cal Z}_{\mbox{MF}}=-i \,
\mbox{Tr} \, \ln \, \left(
i\partial_0-H+\mu\right)
        -\frac{1}{2}\,\int\, d^3x\,B^2 \; ,
\label{eq5}
 \end{equation}
with $H=-D_k^2 /2m$.
Also, the one-loop effective action is given by
\begin{eqnarray}
S = & & \int\,d^3x\,\left( -\frac{1}{4}F_{\mu\nu}F^{\mu\nu}
-\frac{\nu}{2}\epsilon^{\mu\nu\lambda}
   a_\mu \partial_\nu a_\lambda \right) \nonumber \\
   &-& \frac{1}{2}\int\, d^3x\int \,
   d^3y\,\left( A_\mu (x)+a_\mu (x) \right)
   \Pi^{\mu\nu}(x,y)\left( A_\nu (y)+a_\nu (y) \right) \; ,
 \label{eq6}
 \end{eqnarray}
where we have represented the fluctuating fields by $a$ and $A$ again.
The current correlators are given by
\begin{equation}
\Pi^{\mu\nu}(x,y) = -\frac{\delta \langle j^\mu (x)\rangle }
{\delta {\cal A}_\nu (y)}
\Big\vert_{MF}\; ;\;
{\cal A}_\mu =A_\mu +a_\mu
\; ,
 \label{eq7}
 \end{equation}
 The fermionic currents are given by
\begin{mathletters}
\label{eq8}
\begin{eqnarray}
j_0 (x) &=& e\psi^{\dagger}\psi \; , \\
j_k (x) &=& -i\frac{e}{2m}\left( \psi^{\dagger}D_k \psi
     -D_k^\ast\psi^{\dagger}\psi \right)
          \; .
\end{eqnarray}
\end{mathletters}

The single particle Greens function $G(x,y)=-i\langle T \;
\psi (x) \psi^{\dagger}
(y) \rangle $ can be obtained by solving the differential equation
\begin{equation}
\left( i\partial_0 -H+\mu \right) G(x,y)=\delta ^{(3)}(x-y) \; ,
 \label{eq9}
 \end{equation}
subject to appropriate boundary conditions.
The boundary conditions which we use
for evaluating $G(x\, ,\, y)$  will be discussed in the next subsection.
$T$ represents the time ordering of two fermionic fields.
Thus using a suitable limiting procedure one can
express fermionic current
and current correlator respectively in terms of
$G(x\, ,\, y)$ as follows:
\begin{mathletters}
\begin{eqnarray}
\langle j_0 (x) \rangle &=& iG(x,x^\prime )
\Big\vert_{X^\prime =X\, ,\, t^\prime
                     =t+0^+ }   \; ,  \\
\langle j_k (x) \rangle &=& \frac{e}{2m}(D_k-D_k^{\prime\ast}
                     )G(x,x^\prime  )
                  \Big\vert_{X^\prime =X\, ,\, t^\prime =t+0^+} \; ,
\end{eqnarray}
\end{mathletters}
\begin{mathletters}
\begin{eqnarray}
\label{eq11}
\Pi_{00}(x,y) &=& ie^2 G(x,y)G(y,x)    \; ,  \\
\Pi_{0k}(x,y) &=& \frac{e^2}{2m}\left[ G(x,y)D_k^y G(y,x)-\left(
		   D_k^{y\ast}G(x,y)
                 \right) G(y,x) \right] \; ,  \\
\Pi_{kl}(x,y) &=& -i\frac{e^2}{4m^2}\left[ D_k^x G(x,y)D_l^y G(y,x)
                  -\left( D_k^xD_l^{y\ast}G(x,y)\right) G(y,x)
		  \right. \nonumber \\
              & & +D_l^{y\ast}G(x,y)D_k^{x\ast}G(y,x)
	      - G(x,y)D_k^{x\ast}
                  \left.  D_l^{y}G(y,x) \right] \nonumber \\
	      & & -i\frac{e^2}{2m}\delta_{kl}\left(
	      \delta (x-y)+\delta (x^\prime
		     -y)\right) G(x,x^\prime)\Big\vert_{X^\prime =X\, ,
	     \, t^\prime =t+0^+} \; .
\end{eqnarray}
\end{mathletters}

In the MF approximation the current correlators will be obtained
in terms of the Greens functions satisfying (\ref{eq9})
with the MF configuration.
Thus, we pause to discuss the MF ground state first
before discussing the
fluctuations of the gauge fields.

\subsection{Mean Field Results}

The mean field configuration in this case is rather involved
since the external magnetic field changes the degeneracy as well as
the cyclotron frequency. Since the levels are otherwise completely
filled, the highest Landau level (LL) is now only partially filled
which makes the ground state degenerate.

To handle this, we shall introduce a fictitious spin-like
internal degree of freedom, represented by the operator $\hat{U}$
in the Hamiltonian, to the particle which couples to the magnetic field and
splits the degeneracy. (Alternatively, one could also introduce a
background harmonic oscillator potential to split the degeneracy.
However, we find the former choice more convenient).
Indeed, if the degeneracy per unit area is
$\rho_l=1/2\pi l^2$ ($l=\left[ e\vert b+B \vert \right]^{-1/2}$
being the magnetic length of the system), then the `pseudospin'
operator $\hat{U}$ belongs to that representation which has exactly
as many eigenvalues as $\rho_l A$, where $A$ is the area of the
system. Thus, the Hamiltonian
\begin{equation}
H \rightarrow H^\prime =H+\lambda \omega_c \hat{U} \; ,
\end{equation}
where $\omega_c =2\pi\rho_l/m$ is the cyclotron frequency and $\lambda$ is the
dimensionless strength. As mentioned, $\hat{U}$ has eigenvalues
given by
\begin{equation}
-\frac{(\rho_l A-1)}{2} \leq u_i \leq \frac{(\rho_l A-1)}{2} \; ; \;
   u_{i+1}=u_i +1  \; .
\end{equation}
The modified spectrum for $H$ is shown schematically in Fig.~1.
Since $\lambda$ is to be a small parameter (which will be switched
off at the end of our calculation), it is necessary that
\begin{equation}
\lambda \left\vert 1+x \right\vert \ll \frac{N}{\rho A} \; ; \;
  x=\frac{B}{b} \; .
\end{equation}

The spectrum of $H^\prime$ is given by
\begin{equation}
\epsilon_{ni}=\left( n+\frac{1}{2}
 +\lambda u_i \right) \omega_c   \; ; \; n=0,1,2,\cdots
 \; , \; i=1,2,\cdots \rho_l A  \; .
   \end{equation}
The MF Lagrangian (\ref{eq5}) becomes
\begin{equation}
{\cal L}_{MF}=\frac{1}{A}\sum_{n=0}^{\infty}\sum_{i}
\,\int\,\frac{dk_0}{2\pi i}
          \left[ \ln \, (k_0-\epsilon_{ni}+\mu) \right]
	  -\frac{B^2}{2}\, .
\label{eq14}
\end{equation}
In evaluating the density $\rho =\partial {\cal L}_{\mbox{MF}
}/\partial \mu $, we encounter the integral of the form
\begin{equation}
\lim_{\delta \rightarrow 0}\;\int \; \frac{dk_0}{2\pi i}\;
\frac{e^{ik_0 \delta}}{k_0+x+ik_0 \delta}
 = \theta (x)
       \; ,
 \label{eq15}
 \end{equation}
  where an additional convergence term $\exp\,(ik_0 \delta)$
 has been inserted. The integral is thus equal to the heaviside
 function
$\theta (x)= \left\{ \begin{array}{ll} 1 &\mbox{for $x>0$}  \\
0 &\mbox{for $x<0$}  \end{array} \right.  $.
Therefore, expression for the density of fermions follows:
\begin{equation}
\rho \equiv \frac{\partial {\cal L}_{MF}}{\partial \mu}=
\frac{1}{A}\sum_{n=0}^{\infty}\,\sum_i \,
       \theta (\mu -\epsilon_{ni}) \, .
\label{eq16}
\end{equation}
Using the same technique, $k_0$ integral in (\ref{eq14}) can be
determined to obtain
\begin{equation}
{\cal L}_{MF}=\frac{1}{A}\sum_{n=0}^{\infty}\sum_i \,
 (\mu -\epsilon_{ni})\theta (\mu-\epsilon_{ni})-\frac{B^2}{2}
      \; .
\label{eq17}
\end{equation}

Note that the chemical potential (which at $T=0$ equals
the Fermi energy) can be now determined.
In the limit $\lambda\rightarrow 0$,
\begin{equation}
\mu =  \left\{ \begin{array}{l} \left( [K+1]-\frac{1}{2}\right)
\omega_c \\
    K\omega_c  \end{array}  \right.  \; ,
 \label{eq18}
\end{equation}
where $K=N/\vert 1+x \vert $.
The upper case corresponds to fractional filling
factor $K$ and the lower
case to the integer values of $K$.
$[x]$ denotes
the largest integer less than $x$. Note that $\mu$
is discontinuous at those values of $B$,
where it corresponds to integral number
of fully filled LL. In between the two integer
filling fractions $\mu$ varies
linearly with $B$ as $\omega_c$ varies in the same way.
In fact, $\omega_c (x)=\omega_c (0)\vert 1+x \vert $.
Fig.~2 shows the variation of
the chemical potential with the application of
external magnetic field parallel to
the mean CS magnetic field.

The MF Lagrangian for $K$ fractionally filled levels becomes
\begin{equation}
{\cal L}_{MF}=\frac{1}{A}\left( \sum_{n=0}^{[K-1]}\,
\sum_i \, (\mu - \epsilon_{ni}) +
         \sum_{i\leq i_0} \, \left( \mu - \epsilon_{[K]i}
	     \right) \right) -\frac{B^2}{2}
         \; .
 \label{eq19}
\end{equation}
Here $i_0$ denotes the quantum number of the highest occupied
level corresponding to the eigenvalue $u_{i_0}=u_0$ of $\hat{U}$.
The summation over $n$ and $i$ can be easily carried out.
In the limit $\lambda \rightarrow 0$, (\ref{eq19}) gets a simple form
\begin{equation}
{\cal L}_{MF}= \left\{ \begin{array}{l}
\frac{e^2}{4\pi m} \vert b+B \vert^2
	      \left( [K]^2
          +[K]\right) -\frac{B^2}{2}  \\
      \frac{e^2}{4\pi m}\vert b+B\vert^2 K^2 -\frac{B^2}{2} \end{array}
        \right.  \, .
 \label{eq20}
 \end{equation}

The last term in Eq.(\ref{eq20}) is due to the
kinetic term of the Maxwell
field and  does not contribute in the calculation of electro-magnetic
response.
The MF magnetic susceptibility can be readily
evaluated (by the omission
of Maxwell kinetic term) as
\begin{equation}
\chi_{MF}\equiv\frac{\partial^2 {\cal L}_{MF}}{\partial B^2}=
  \left\{ \begin{array}{l} \frac{e^2}{2\pi m}\left( [K]^2 +[K] \right)
          \\  \frac{e^2}{2\pi m}K^2  \end{array} \right.  \; .
\label{eq21}
\end{equation}
This is also discontinuous when $[K]$ passes
from one value to the other.
This is the well known de Haas-van Alphen effect
for diamagnetism.
Recall that susceptibility diverges quadratically for $B \rightarrow 0$
in the conventional de
Haas-van Alphen effect where there is no internal field.
\cite{huang} Similarly, here, it diverges for $B\rightarrow
-b$.
The variation of $\chi_{MF}$ with the magnetic field is
shown in Fig.~3.
We note that a similar behaviour was also obtained
by Hetric  et al. \cite{het1}
in their calculation of MF magnetization, which we obtain as
\begin{equation}
{\cal M}_{MF}=-\frac{e\rho}{2m}\left( 1+2[K]-2\frac{[K]}{N}\left(
 [K]+1 \right) \left\vert 1+\frac{B}{b}\right\vert \right) \mbox{sign}(b+B) \;
{}.
 \end{equation}
However, they do not consider the fluctuations of the gauge fields
which should play an important role
in the response of the system to
the applied magnetic field.
We shall pursue this issue in the
next section.

\subsection{Current Correlation Functions}

The evaluation of the form factors requires the Green function
(\ref{eq9}). We first define the frequency transformed Greens function
$G_\omega (\vec{X}\, ,\, \vec{X}^\prime)$ to be
\begin{equation}
G(x,x^\prime)=\int\,\frac{d\omega}{2\pi}e^{-i\omega (t-t^\prime)}
G_\omega (\vec{X}\, ,\, \vec{X}^\prime) \; ,
\end{equation}
which clearly satisfies the differential equation
\begin{equation}
\left[ \omega -H^\prime +\mu \right]
 G_\omega (\vec{X}\, ,\, \vec{X}^\prime ) =\delta^{(2)}
 (\vec{X} - \vec{X}^\prime ) \; .
 \end{equation}
 This can be solved, as usual, in terms of the complete set of
 eigen functions $\psi_{ni}(\vec{X})$ of $H^\prime $ giving
 \begin{equation}
 G(x\, ,\, x^\prime)=\int_C \,\frac{d\omega}{2\pi}\,\sum_{ni}
 \frac{\psi_{ni}(\vec{X})\psi_{ni}^\ast (\vec{X}^\prime)}{\omega
 -\epsilon_{ni}+\mu } e^{-i\omega (t-t^\prime)} \; .
 \end{equation}
 The contour $C$ for the frequency integration has to be chosen so that
 $G(x\, ,\, x^\prime)$ satisfies the boundary conditions
\begin{equation}
G(x\, ,\, x^\prime )\sim \left\{ \begin{array}{ll}
\sum_{n =0}^{[K-1]}\sum_i \,
e^{-i(\epsilon_{ni}-\mu )(t-t^\prime)}+\sum_{i\leq i_0}\,
e^{-i(\epsilon_{[K]i}-\mu )(t-t^\prime )}\, , & t<t^\prime \, , \\
\sum_{n >[K]}\sum_i \, e^{-i(\epsilon_{ni}-\mu )
(t-t^\prime)}+\sum_{i >i_0}\,
e^{-i(\epsilon_{[K]i}-\mu )(t-t^\prime )}\, , & t>t^\prime \: .
\end{array}    \right.
\label{eq22}
\end{equation}
More explicitly, the contour $C$ must pass below the poles at
$\omega =\epsilon_{ni}-\mu$ for $n\leq [K-1]$, for all $i$ and $n=[K]$,
$i\leq i_0$ and above otherwise.

We now evaluate current correlation functions in the momentum
space following the elegant procedure
given by Randjbar-Daemi  et al.
\cite{salam}
Gauge and rotational invariance requires
that the current correlator has the form
\begin{eqnarray}
\Pi^{\mu\nu} (\omega\; , \; \vec{q}^2 )
=& &\Pi_0 (\omega\; , \; \vec{q}^2 )\;
 (q^2 g^{\mu\nu}
          -q^\mu q^\nu ) \nonumber \\
       &+&(\Pi_2 -\Pi_0 )\; (\omega \; , \; \vec{q}^2 ) \delta^{\mu i}
           \delta^{\nu j}
   (\vec{q}^2 \delta^{ij} -q^i q^j )+i\Pi_1 (\omega \; , \; \vec{q}^2
              ) \epsilon^{\mu \nu \lambda}
             q_\lambda \, ,
\label{eq24}
\end{eqnarray}
where the last term is the parity and time reversal
violating contribution. Here $q^2=\omega^2 -\vec{q}^2 $.

For the purpose of evaluating low energy
effective Lagrangian, it is sufficient
to compute the form factors $\Pi_0$, $\Pi_1$
and $\Pi_2$ in the limit $\omega
\rightarrow 0$, $\vec{q}^2 \rightarrow 0$.
The limits commute in the evaluation of form factors.
For simplicity, we take the limit $\vec{q}^2 \rightarrow 0$ first.
 In the limit $\vec{q}^2 \rightarrow 0$, the form factors are given
by
\begin{mathletters}
\label{eq25}
\begin{eqnarray}
\Pi_0 (\omega \, , \, 0) &=& \frac{e^2}{2\pi \rho_l A}\int\,\frac{d
 \omega^\prime}{2\pi i}\sum_{ni} \,
\sum_{m j} \frac{\left( n \delta_{n,\, m+1}
+(n+1)\delta_{n,\, m-1}\right)
 \delta_{ij}}{(\omega^\prime -\epsilon_{mj}+\mu )
 (\omega^\prime -\epsilon_{ni}
                   +\mu -\omega )}   \; ,  \\
\Pi_1 (\omega \, , \, 0) &=& \Pi_0 \omega_c  \; , \\
\Pi_2 (\omega \, , \, 0) &=& \frac{e^2\omega_c}{2\pi m\rho_l A} \int\,
       \frac{d\omega^\prime}{2\pi i} \sum_{ni} \, \sum_{mj}\frac{1}{
       (\omega^\prime -\epsilon_{mj}+\mu )(\omega^\prime -\epsilon_{ni}
 +\mu -\omega)} \nonumber \\
 & & \times \left[  n(n-1)\delta_{n,\,m+2}
        +3n^2\delta_{n,\, m+1}
        +(2n+1)^2\delta_{nm} \right. \nonumber \\
	& & +\left. 2(n+1)^2\delta_{n,\, m-1}+(n+1)(n+2)
       \delta_{n,\, m-2} \right] \delta_{ij} \; .
\end{eqnarray}
\end{mathletters}
The integral which we encounter in
the evaluation of (\ref{eq25}) is
\begin{eqnarray}
 \int\; \frac{d\omega^\prime}{2\pi i} & & \frac{1}{(\omega^\prime
  -\epsilon_{mj} +\mu )(\omega^\prime
       -\epsilon_{n i} +\mu -\omega )} \nonumber \\
       &=& (\epsilon_{ni}-\epsilon_{mj} +\omega )^{-1} \; ,
       \hspace{0.1in}\mbox{for}\left\{
 \begin{array}{ll}
       m \geq [K+1]\; ,\; \mbox{for all $j$}\, ;\, & m=[K]\; ,\; j>i_0\, ;  \\
       n<[K]\; ,\;\mbox{for all $i$}\, ; &  n=[K]\; ,\; i\leq i_0
 \end{array}   \right.
 \label{eq26}  \\
          &=& -(\epsilon_{ni}-\epsilon_{mj} +\omega )^{-1} \, ;
           \hspace{0.15in} \mbox{for}\left\{
 \begin{array}{ll}
       n \geq [K+1]\; ,\; \mbox{for all $i$}\, ;\, & n=[K]\; ,\; i>i_0\, ;  \\
       m<[K]\; ,\;\mbox{for all $j$} \, ; & m=[K]\; ,\; j\leq i_0
 \end{array}   \right.
         \nonumber \\
         &=& 0 \hspace{1.7in} \mbox{otherwise}
         \, . \nonumber
 \end{eqnarray}
Thus, the form factors (in the limit
$\omega \rightarrow 0\; ,
 \; \vec{q}^2
\rightarrow 0$,) are obtained as
\begin{equation}
\Pi_0=\frac{e^2K}{2\pi \omega_c}\, , \, \Pi_1=\frac{e^2K}{2\pi}\, , \,
\Pi_2=\frac{e^2K^2}{2\pi m}    \, .
 \label{eq27}
\end{equation}
Note that they are dependent on the applied magnetic field.
In terms of their values at $B=0$, they are expressed as follows:
\begin{mathletters}
\label{eq28}
\begin{eqnarray}
\Pi_0 (x)&=& \frac{e^2mN^2}{4\pi^2\rho\left|
1+x \right| ^2}=\frac{\Pi_0 (0)}{
\left| 1+x \right| ^2 }\, , \,  \\
\Pi_1 (x)&=& \frac{e^2N}{2\pi\left| 1+x \right| }=\frac{\Pi_1 (0)}{
\left| 1+x \right| }\, ,\, \\
\Pi_2 (x)&=& \frac{e^2N^2}{2\pi m \left| 1+x
\right| ^2} = \frac{\Pi_2 (0)}{
\left| 1+x \right| ^2   }\; .
\end{eqnarray}
\end{mathletters}
The behaviour of the form factors for negative $x$, specially for
$x=-1$, will be discussed later in section (IV).

\section{{\bf GAUGE FIELD FLUCTUATIONS}}

\subsection{{\bf Effective Lagrangian}}

Now we choose the Coulomb gauge,
$\partial_i A_i =0$; $\partial_i a_i =0$
in order to evaluate the fluctuating part of the
partition function (\ref{eq4})
by an integration over the gauge fields. Basically one computes the
contribution to the partition function from the effective gauge field
modes (collective excitations). We obtain
\begin{equation}
 \ln \, {\cal Z}_{\mbox{eff}}=\frac{i}{2}
 \, \mbox{Tr}\, \ln \, \left[ \left(
      \omega^2 -\omega_+^2 \right) \left( \omega^2 -\omega_-^2 \right)
	 \right]  \; ,
 \label{eq29}
\end{equation}
The collective modes have the dispersion relation $\omega =
\omega_\pm (\vec{q}^2)$, with
\begin{equation}
\omega_\pm^2 = \frac{1}{2C_1}\left(C_2 \pm
\sqrt{C_2^2-4C_1C_2} \right)  \; ,
\label{eq30}
\end{equation}
and
\begin{mathletters}
\label{eq31}
\begin{eqnarray}
C_1 &=& \Pi_0^2  \; ,\\
C_2 &=& \Pi_0 (\Pi_0 +\Pi_2 ){\vec q}^2 +\nu^2 \left(
\Pi_0^2+2\Pi_0)+(\nu
	-\Pi_1\right) ^2 \; , \\
C_3 &=& \Pi_0\Pi_2 {\vec q}^4 +\left[
\nu^2 \left(\Pi_0 +\Pi_2+\Pi_0\Pi_2\right)
	 +(\nu -\Pi_1)^2 \right] {\vec q}^2 +\nu^2\Pi_1^2  \; .
\end{eqnarray}
\end{mathletters}
Therefore, one finds the effective Lagrangian
\begin{equation}
{\cal L}_{\mbox{eff}}=\frac{i}{2}\int_{-\infty}^{\infty}
    \, \frac{d\omega}{2\pi}
\int\, \frac{d^2q}{(2\pi)^2}\,
\ln \, \left[ (\omega^2-\omega_+^2)(\omega^2-\omega_-^2) \right]  \; .
\label{eq32}
\end{equation}
The divergent frequency integral can be regularized as in Eq.~(\ref{eq16})
to obtain
\begin{equation}
{\cal L}_{\mbox{eff}}=-\frac{1}{2}\int\,
\frac{d^2q}{(2\pi)^2}\, (\omega_+
  +\omega_-)  \; .
\label{eq33}
\end{equation}
This is the effective Lagrangian for the system coming from the
fluctuations of the gauge fields, where the dependence on the external
magnetic field arises through the form factors in the dispersion relations.
Below we discuss the behaviour of $\omega_\pm$ as functions of the
magnetic field.

\subsection{Low-lying Excitations}

We first consider, for simplicity,
the neutral system (i.e., excluding
the internal Maxwell gauge field).
The dispersion relation that we obtain, for the
neutral system, from the equation of motion
of the CS gauge field is given by
\begin{equation}
\Pi_0\,(\Pi_0\omega^2-\Pi_2{\vec q}^2)-(\nu -\Pi_1)^2 =0 \; .
\label{eq34}
\end{equation}
Recall that in the absence of an external magnetic field,
the tree level CS term exactly
canceled with the dynamically generated CS term
(i.e., $  \nu = \Pi_1$). Thus, the
system possessed a super-fluid mode with the low-lying massless
phononic excitation,
\begin{equation}
\omega^2=\frac{\Pi_2(0)}{\Pi_0(0)}{\vec q}^2   \; .
\end{equation}
(This in fact is the collective
pseudo-Goldstone mode which is absorbed by
the Maxwell gauge field upon coupling
the particles to the EM field
thereby producing a gap and leading to the Meissner effect).

When we introduce external magnetic field,
this mode acquires a mass $(M)$,
\begin{equation}
M^2=\omega_c^2(0)x^2(1+x)^2 \; ,
\end{equation}
as the cancellation of the CS term no longer holds
and hence the super-fluidity diminishes. The dispersions are shown in
Fig.~4 for three different values of $B$.
We have chosen \cite{ssman1,het1}
$N=10$, $e^2=10^5 \mbox{cm}^{-1}$ and the value of two dimensionless
parameters $e^2/m $ and $ \rho /me^2 $ to be $10^{-5}$ and $10^{-1}$
respectively.
Observe that the mass gap increases with
the value of $B$ although the dispersions are parallel. Thus, the
velocity of the mode decreases with the increase of $B$.

We now come back for a discussion of excitations of the charged system.
The low-lying excitations of the system are
two massive photonic modes. These two
modes are nothing but $\omega_\pm$ given in equations
(\ref{eq30} and \ref{eq31}).
Fig.~5 and Fig.~6 show how the frequencies
disperse with momentum for three chosen values of $B$ for the
modes. The masses $M_\pm^2$ increase with $B$. Note however, that the
damping lengths $\lambda_\pm^2=M_\pm^{-2}$ have to be distinguished from
the penetration depth, \cite{ssman1,het1} which in fact decreases with
increasing $B$, signifying the absence of the Meissner effect. This is
seen from the calculation of the magnetic susceptibility below.

\section{Nonlinear response}

Here we study the nonlinear response of the system to an external field by
computing the higher order susceptibilities defined by
\begin{equation}
\chi^{(r)}\equiv -\left. \frac{1}{(r+1)!}\frac{\partial^{r+2}F}{\partial
 B^{r+2}} \right\vert_{B=0} \;
;\; r=0,1,2,\cdots  \; ,
\label{eqr1}
\end{equation}
where $F$ is the free energy density of the system which
can be computed from the
Lagrangian (\ref{eq33}) as $F\equiv -{\cal L}_{\mbox{eff}}$
and $\chi^{(0)}$ is recognized
to be the linear response susceptibility. Note that the complete response
of the system is obtained by adding the
corresponding MF values which may be inferred from Eq. (\ref{eq21}).

We expand the sum $\omega_++\omega_-$ up to
order ${\vec q}^2$, since it is
a low energy-momentum theory, and obtain
\begin{equation}
\omega_++\omega_- = F_1 + F_2 {\vec q}^2
\end{equation}
with
\begin{mathletters}
\begin{eqnarray}
F_1 &=& \nu \left[ 1+\frac{2}{\Pi_0}+\frac{1}{\Pi_0^2}
	\left( 1-\frac{\Pi_1}{\nu}\right) ^2
	  +\frac{2\Pi_1}{\Pi_0 \nu}  \right]^{\frac{1}{2}}   \; , \\
F_2 &=& \frac{1}{2F_1}\left[ \left( 1+\frac{\Pi_2}{\Pi_0}\right)
	 +\frac{\nu}{\Pi_1}\left(
	 \Pi_0+\Pi_2+\Pi_0\Pi_2+\left( 1-\frac{\Pi_1}{\nu}\right) ^2
	 \right) \right] \; .
\end{eqnarray}
\end{mathletters}
We introduce a momentum cut-off $\Lambda$ to obtain
\begin{equation}
F=\frac{1}{8\pi}\left[ F_1\Lambda^2 +\frac{1}{2}F_2
	  \Lambda^4 \right] \; .
\label{eq39}
\end{equation}

The susceptibilities $\chi^{(r)}$, which are apparently a function
of $\Lambda$, are
calculated to be
\begin{equation}
\chi^{(r)}(B,\Lambda)
   = -\frac{\Lambda^2}{8\pi (r+1)!}\left[ \frac{\partial^{r+2}
   F_1}{\partial B^{r+2}}
	    +\frac{\Lambda^2}{2}\frac{\partial^{r+2}F_2}{\partial B^{r+2}}
	    \right]_{B=0}  \; .
\label{eq40}
\end{equation}
To determine $\Lambda$, we demand that the
linear magnetic susceptibility
\begin{equation}
\chi^{(0)}\equiv -\left. \frac{\partial^2F}{ \partial B^2}
\right\vert_{B=0} =-1 \;,
\end{equation}
which we know already
and independently from
linear response theory \cite{ssman1,ssman2}.
This fixes $\Lambda^2$ to be
\begin{equation}
\Lambda^2 \approx \frac{8\pi^2\rho}{N}\left[
\left( \frac{2\rho}{e^2mN}\right)
	   ^{\frac{1}{2}}
	   -\frac{\rho}{m^2N} \right]  \; .
\label{eq41}
\end{equation}
This value of $\Lambda$ is appropriate as a
cut-off momentum for high magnetic field also
since $\Lambda^2l^2 \approx 1.8/\vert 1+x \vert$, using our chosen values
of parameters. Therefore,
unless $x$ is very high, $\Lambda$ is quite reasonable as a
cut-off momentum.
The cut-off independent non-linear magnetic susceptibilities
can now be readily evaluated from (\ref{eq40}) by the substitution
of the value of $\Lambda^2$ (\ref{eq41}).
They can be analytically determined, to a high degree of approximation to be
\begin{equation}
\chi^{(r)}=(-1)^{r+1}\frac{r+2}{2b^r}  \; .
\label{eqr2}
\end{equation}
Notice that higher order susceptibilities are nonvanishing and do not even
have numerically small values,
which clearly shows that there is neither a critical nor a
pseudo critical field which would characterize the phase of the system.
However, as $B\rightarrow \infty$, the free energy
has only a linear dependence on $B$,
which means that all the susceptibilities vanish.
The system returns to its normal state asymptotically. We also notice
that since both even and odd order susceptibilities survive, the system is
magnetically asymmetric around $B=0$, as a consequence of the ${\cal P}$,
${\cal T}$ violation inherent in the theory.

It is instructive to study the behaviour of free energy as a function of
applied magnetic field. It has the (approximate) form
\begin{equation}
F(B)\approx \frac{b^2}{2(1+B/b) }-\frac{b^2}{2}(1-B/b) \; .
\label{eqr3}
\end{equation}
Clearly, the leading order contribution comes from $B^2$ since,
 as we know from the linear response analysis, \cite{ssman1,ssman2} the
system does not possess any spontaneous magnetization.
The curvature of $F(B)$ at $B=0$ gives the linear susceptibility.
The behaviour
of $F(B)$ is shown in Fig.~7. It may be seen (see also Eq. (\ref{eqr3}))
that the free energy diverges at $B=-b$. This requires
some discussion, as it corresponds to a zero mean field.

As $B\rightarrow -b$, the inter Landau level
spacings squeeze and hence the energy spectrum approaches the
continuum; more and more number of LL will be filled up even as the
particle density remains the same. Therefore, the values of the
form factors (\ref{eq28}) increase and they diverge leading to the
divergence in $F$. The correctness of the above observations hinges
crucially on
the validity of the MF ansatz in this region.
Some indirect support for the validity can
be obtained from the related phenomena of FQHE, where the
composite fermion model \cite{jain} predicts a zero mean field at
the  filling
fraction $\nu =1/2$, thus leading to a free fermion like system. The analysis
of Halperin, Lee and Read \cite{halp} and a recent experiment
by Du et al. \cite{tsui} do support this prediction. More specifically
for the system at hand, we recall the argument given in Ref.~6.
One simple criterion is that MF would
be plausible if the interparticle seperation is less than the magnetic
length $l$ which is a measure of the scale over which the single
particle wave function extends,  i.e.,
\begin{equation}
\sqrt{\rho^{-1}} \;\widetilde{<}
l \; \Rightarrow \; \frac{N}{\vert 1+x \vert}
\widetilde{>} 2\pi \; .
\label{eq41f}
\end{equation}
Therefore, the MF theory will be invalid for very high $\vert x\vert $.
However for moderate $\vert x\vert $, the MF ansatz works very well.
Note that (\ref{eq41f}) suggests that the MF theory is valid at $x=-1$.

Before we end this section, we remark that the MF value for $F$
approaches a constant value, with an ever increasing number of discontinuities
as $x\rightarrow -1$. In
contrast, the contribution from the fluctuations diverges linearly and thus
dominates over the MF contribution.
To illustrate this, if the  generalized susceptibility defined as
$\chi (x)\equiv \partial {\cal M} (x)/\partial B$ is
considered, it is easy to see that the fluctuation part diverges
as $-1/(1+x)^3$ unlike the MF contribution which diverges only
quadratically in the same limit. Finally, note that
even around $x=0$, the susceptibilities get a dominant
contribution from the fluctuations.

\section{conclusion}

To conclude, the mean field properties such
as chemical potential, magnetization and magnetic susceptibility oscillate
as functions of the external magnetic field. Considering the
fluctuations of the gauge fields we find that the masses of the mode of
excitations increase with magnetic field. The nonlinear susceptibilities
arising from the fluctuations of the gauge fields are computed which show
the absence of any critical or pseudo critical field. Further,
there is a unique asymmetry in the system at $B=0$ as far as the magnetic
properties are concerned, and would be an interesting property to look for
in case candidates for CSS are proposed in real systems.

\newpage

\begin{figure}
\caption{The energy levels for MF ground state. $N$ Landau Levels are
    filled at $B=0$. At $B\neq 0$, there are $K$ fractionally
    filled (topmost level is not fully filled) levels. Split levels
    arise due to the switching on of the fictitious interaction.
    Each split level can accomodate only one particle. The number of
    split levels for each LL are equal to the degeneracy of each LL.
    The horizontal dotted lines represent the Fermi levels.
    }
\end{figure}

\begin{figure}
\caption{Chemical potential $\mu$ (solid line)
is plotted against the applied
	  magnetic field $B$, $N=6$, in the units of $2\pi \rho /m
	  =\mu (B=0)$. $B$
  is in the units of CS magnetic field $b$. Note that $\mu$ shows
  discontinuities at those values of $B$ where the number of filled
  levels are exactly integral in number. Also, note that the values of
	  $\mu$ at integer filling (denoted by dark dots) are the
	  same and equal that at $B=0$.}
\end{figure}

\begin{figure}
\caption{MF magnetic susceptibility $\chi_{\mbox{MF}}$ (solid lines)
    is shown, for $N=6$, against $B/b$.
    It also shows
 the discontinuities in those values of $B$ for which filled levels
 are integers. Its values for those values of $B$ are shown by
	 dark dots.}
\end{figure}

\begin{figure}
\caption{The dispersions of the phononic
modes for different values of $B$
 are shown. Here $l_0=\left| eb \right|^{-1/2}$ is the magnetic length
 of the system in absence of $B$. The  numbers associated to each curve
 are the applied magnetic field $B$ in units of $b$. Note that
	 the mass gap of the mode increases with $B$.}
\end{figure}

\begin{figure}
\caption{The dispersion relation
$\omega_+^2$ as a function of $\vec{q}^2$ for different values of $B$.
	 The numbers associated to each curve are the applied
	 magnetic field $B$ in units of $b$.}
\end{figure}

\begin{figure}
\caption{The dispersion relation $\omega_-^2$
for different values of $B$.
	  The numbers associated to each curve are the applied
	 magnetic field $B$ in units of $b$.}
\end{figure}

\begin{figure}
\caption{Free energy is shown as a function of magnetic field.
	 }
\end{figure}

\end{document}